\documentclass[10pt,journal,compsoc]{IEEEtran}

\usepackage[switch]{lineno}
\AtBeginDocument{%
  \providecommand\BibTeX{{%
    \normalfont B\kern-0.5em{\scshape i\kern-0.25em b}\kern-0.8em\TeX}}}

\usepackage{url}
 \usepackage{graphicx}
\usepackage[inline]{enumitem}
\usepackage{subfig}
 \usepackage{tikz}
 \usepackage{makecell}
 \usepackage{algorithm}
 \usepackage{algorithmic}
\usetikzlibrary{arrows,positioning,shapes.geometric}
\usepackage{tcolorbox}
\usepackage[multiple]{footmisc}
\begin{document}
\title{\Large \bf \textit{Trusting code in the wild}: A social network-based \\centrality rating for developers in the Rust ecosystem}

\author{Nasif~Imtiaz, Preya Shabrina,
        and~Laurie~Williams% <-this % stops a space
\IEEEcompsocitemizethanks{\IEEEcompsocthanksitem Nasif Imtiaz, Preya Shabrina, and Laurie Williams are with the Department of Computer Science, North Carolina State University, Raleigh, NC, 27606.\protect\\
% note need leading \protect in front of \\ to get a newline within \thanks as
% \\ is fragile and will error, could use \hfil\break instead.
E-mail: {simtiaz, pshabri, lawilli3}@ncsu.edu}% <-this % stops an unwanted space
% \thanks{Manuscript received April 19, 2005; revised August 26, 2015.}
}

% \author{Nasif Imtiaz}
% \affiliation{%
%  \institution{North Carolina State University}
%  \city{Raleigh}
%   \state{North Carolina}
%  \country{U.S.A}
% }
% \author{Preya Shabrina}
% \affiliation{%
%  \institution{North Carolina State University}
%  \city{Raleigh}
%   \state{North Carolina}
%  \country{U.S.A}
% }
% \author{Laurie Williams}
% \affiliation{%
%  \institution{North Carolina State University}
%  \city{Raleigh}
%   \state{North Carolina}
%  \country{U.S.A}
% }
\IEEEtitleabstractindextext{
\begin{abstract}

As modern software extensively uses open source packages, developers regularly pull in new upstream code through frequent updates. While a manual review of all upstream changes may not be practical, developers may rely on the authors' and reviewers' identities, among other factors, to decide what level of review the new code may require. The goal of this study is to help downstream project developers prioritize review efforts for upstream code by providing a social network-based centrality rating for the authors and reviewers of that code. To that end, we build a social network of 6,949 developers across the collaboration activity from 1,644 Rust packages. We then compute a rating for each developer based on five centrality measures extracted from the network. Further, we survey the developers in the network to evaluate if code coming from a developer with a higher centrality rating is likely to be accepted with lesser scrutiny by the downstream projects and, therefore, is perceived to be more trusted.

Our results show that 97.7\% of the developers from the studied packages are interconnected via collaboration, with each developer separated from another via only four other developers in the network. The interconnection among developers from different Rust packages establishes the ground for identifying the central developers in the ecosystem. Our survey responses ($N=206$) show that the respondents are more likely to not differentiate between developers in deciding how to review upstream changes (60.2\% of the time). However, when they do differentiate, our statistical analysis showed a significant correlation between developers' centrality ratings and the level of scrutiny their code might face from the downstream projects, as indicated by the respondents. Our findings indicate that social network-based centrality rating can be used to estimate the trustworthiness of a developer at a package ecosystem level and, therefore, may help downstream projects decide what level of review new upstream changes require.

\end{abstract}
}
\maketitle

\section{Introduction}

Modern software extensively uses open source packages in its supply chain~\cite{blackduck2021}. However, the use of open source packages has opened up new attack vectors, as vulnerable and even malicious code can sneak into software through these packages~\cite{ohm2020backstabber}. Therefore, researchers and practitioners have recommended security and quality control measures on the clients' end before accepting new package code, such as provenance verification and checking for weak links or potential malware in the package~\cite{slsa, zahan2022weak, torres2019toto, ferreira2021containing, vu2021lastpymile, ferreira2021containing}.

Imtiaz and Williams ~\cite{imtiaz2022phantom} have proposed another quality control method that measures code review coverage in a package update to determine what portion of changes in the update has been reviewed by a second developer other than the author.
Further, SLSA (Supply chain Levels for Software Artifacts), a security framework for using open source packages, also suggests a two-person code review requirement as an industry best practice~\cite{slsa}.
Software practitioners have developed tooling so that third parties can review code in already published packages, such as \textit{cargo-crev}~\cite{cargocrev} and \textit{cargo-vet}~\cite{cargovet}.
However, the mere presence of a review approval in the open source world may not ensure that the involved developers, both authors, and reviewers, are trustworthy. 

A user account management system can help authenticate the users and ensure their trustworthiness within an organizational context. However, the open source world has no centralized account management system~\cite{hawthorne2005software}. Brewer et al.~\cite{googleopensource} noted that a lack of user authentication might make code reviews in open source unreliable,
as developer accounts on online coding platforms such as GitHub cannot be assumed to be independent or trustworthy. They called for a federated model for developer digital identities~\cite{googleopensource}.

However, to the best of our knowledge, we do not know of any such identity model for open source developers to date. A rating or ranking attributed to each developer that reflects their standings within a community can help establish such an identity model. In this work, we conjecture that the public development history of open source packages can be leveraged to build such a rating system. We propose constructing a collaboration graph of the developers using the development history, referred to as developer social network (DSN) in the literature~\cite{herbold2021systematic}. We can then rate the developers based on their positional properties in the graph. The rating (or ranking within the community) can help understand the trustworthiness of open source developers. For example, code changes that are authored and reviewed by top-ranked developers may be more trustworthy. On the other hand, changes coming from newcomers or fringe developers should be reviewed more carefully by the downstream projects. 

The goal of this study is to help downstream project developers prioritize review efforts for upstream code by providing a social network-based centrality rating for the authors and reviewers of that code. To that end, we study the Rust ecosystem in this paper —that is— the packages hosted on Crates.io~\cite{cratesio}. We conduct our study by investigating the following research questions:

\begin{quote}
    \textbf{RQ1:} To what extent are developers from the most downloaded Rust packages interconnected through collaboration? 
\end{quote}

Towards RQ1, we construct a social network of developers from the most downloaded Rust packages based on their collaboration activity within these packages, such as author-reviewer collaboration, and investigate how interconnected the developers from the different packages are. If the developers are interconnected, a social network analysis can provide meaningful signals for an individual developer's position in the network~\cite{sherchan2013survey}.

\begin{quote}
   \textbf{RQ2:} 
   What is the distribution of the developer ratings based on their network centrality?
\end{quote}

Towards RQ2, we propose rating developers using an aggregation of five centrality measures computed from our constructed network. Prior work in social networks, including developer social networks, has demonstrated the use of centrality measures to quantify the influence, trust, social capital, and reputation of individual actors in a network~\cite{bosu2014impact, asim2019trust}. Specifically, centrality measures have been used in various social network-based trust models~\cite{asim2019trust, ceolin2017social, meo2017using, zahi2020improved, csimcsek2020combined} where trust is defined as ``a measure of confidence that an entity will behave in an expected manner, despite the lack of ability to monitor or control the environment in which it operates''~\cite{sherchan2013survey,singh2007privacy}. We hypothesize that a centrality-based rating system can help identify the developers whose code may be perceived as more trusted by the community and, therefore, be accepted by the downstream projects with less scrutiny.

\begin{quote}
    \textbf{RQ3:} Are developers with higher centrality ratings likely to have their code accepted with less scrutiny by the downstream project developers?
\end{quote}

Towards RQ3, we survey the Rust developers in our constructed network to understand if the authors' and reviewers' identities impact how carefully they review upstream code. We designed our survey questions to evaluate if our proposed rating correlates with developers' perception of the trustworthiness of another developer, in terms of code contributed by them getting accepted by downstream project developers with less scrutiny.

Overall, we make the following contributions in this paper:
\begin{enumerate}

    \item A social network-based rating of Rust developers derived from an aggregation of five centrality measures;
    \item A survey-based evaluation of our rating approach in terms of a correlation between a developer's centrality rating and the level of review their code may face from the downstream project developers;
    \item Empirical insights on the developer community structure of Rust ecosystem, highlighting the interconnected nature of developers from different packages;
    \item A dataset on collaboration activity among developers in the Rust package ecosystem.
\end{enumerate}

The Institutional Review Board has approved our study design (\textit{IRB}). Our dataset will be published at \url{https://figshare.com/s/704b51b462e57c97e2fd}.
The rest of the paper is structured as follows: 
Section~\ref{relwork} discusses the background and related work to this study. Section~\ref{sec:method} explains our methodology to construct a social network of Rust developers and compute a centrality rating for each developer. We also present our survey questionnaire in this section, where we seek to understand if developers with higher centrality ratings are more trusted by the community. Afterward, Section~\ref{sec:rq1}, \ref{sec:rq2}, and \ref{sec:rq3} present our findings for the three research questions in this study. Section~\ref{sec:discussion} and \ref{limitation} discusses the implications and limitations of our study before we conclude this paper in Section~\ref{conclusion}.

\section{Background \& Related Work}
\label{relwork}

In this section, we explain the key concepts of our study and discuss the related work. 

\subsection{Package Supply Chain}
Modern software extensively uses open source packages~\cite{blackduck2021}. 
Each major programming language has a package registry supplying a large number of freely available packages, e.g., npm for JavaScript, PyPI for Python, Crates.io for Rust, and RubyGems for Ruby. 
Developers writing software in these languages can use these packages and have the benefit of code reuse~\cite{feitosa2020code}. When projects include open source packages in their codebase, the package becomes an upstream \textbf{\textit{dependency}}. 
Practitioners refer to these packages as a part of the supply chain of the client software~\cite{supplychain}. 
%In this paper, we study the packages hosted on Crates.io that constitute the open source supply chain for software written in Rust. 

\subsection{Emerging Industry Standards for Dependency Packages}
While they provide the benefit of code reuse, open source packages also come with security risks. The risk of known and unknown vulnerabilities in the dependencies has been studied in the literature~\cite{decan2018impact, lauinger2018thou, imtiaz2022open, imtiaz2021comparative}. 
Another risk of using open source packages is the possibility of supply chain attacks~\cite{ohm2020backstabber}. Attackers have compromised popular packages and used the package to conduct malicious attacks~\cite{ohm2020backstabber}. Therefore, practitioners now recommend reviewing the package code before adding or updating a dependency~\cite{imtiaz2022open,yang2021solarwinds}. Consequently, industry standards have emerged that provide a checklist of controls for the safe use of open source packages. A Notable example is Supply Chain Levels for Software Artifacts (SLSA)~\cite{slsa}.

Prior research has shown that developers may decide on accepting new package code based on how trustworthy the maintainers appear to be~\cite{zahan2022weak, wermke2022committed}. Similarly, SLSA has put emphasis on human evaluation of code by requiring open source code to be two-person reviewed~\cite{slsa}. 
Following SLSA requirements, Imtiaz and Williams ~\cite{imtiaz2022phantom} have proposed measuring code review coverage (\textit{CRC}) when accepting dependency updates. However, they note that without ensuring the trustworthiness of the authors and reviewers, the \textit{CRC} metric by itself may not be adequate. Existing work thus shows the need for an identity model for open source developers that may reflect the developer's reputation and past activity history within a particular ecosystem. We address this need in this paper through an approach of a social network-based developer ranking system.

\subsection{Developer Social Network (DSN)}
Social network analysis is the process of analyzing social structures through the use of network and graph theory. The network conceptualizes individuals as nodes and their relations as edges~\cite{scott2012social}. In a developer social network (DSN), the nodes are developers, and the edges represent their communication or collaboration relations~\cite{herbold2021systematic}. A rich body of literature exists that studies DSN~\cite{herbold2021systematic, schreiber2020social, bosu2014impact, meneely2011socio}.

Herbold et al.~\cite{herbold2021systematic} have conducted a systematic mapping study of DSN research. They find nearly half of the research investigates the structure of the developer community. Further, they find that most DSN studies work with a small sample of projects. 
% Besides community structure, research has used DSN data to predict defects~\cite{abreu2009developer}, project outcome~\cite{cataldo2012impact}, build failure~\cite{schroter2010predicting}, post-release failure~\cite{meneely2008predicting}, and understand the collaboration behavior among the developers~\cite{cohen2018large, feczak2009measuring,gharehyazie2017tracing}.
Meneely et al.~\cite{meneely2011socio} have investigated if the metrics obtained from sociotechnical developer networks are reliable through a developer survey. They found statistical evidence for DSN metrics to represent how developers collaborate in practice. Further, Nia et al. ~\cite{nia2010validity} found the network measures to be robust even with incomplete data or missing links between some developers.

Prior research on DSN has primarily focused on single large-scale projects~\cite{herbold2021systematic}. However, open source packages are typically split based on specific functionalities, and the developers may collaborate on many related packages. Research has shown the complex dependency network between packages within an ecosystem~\cite{kikas2017structure}. Therefore, we conjecture that developers in the Rust supply chain may be interconnected as well. Consequently, we may leverage the developer network at a package ecosystem level and measure social metrics like trust and reputation, as was done in prior DSN work focused on single large-scale projects~\cite{meneely2011socio, bosu2014impact}.

\subsection{Social Network Measures}
In this subsection, we explain common network analysis methods and centrality measures.  

\subsubsection{Network Structures} 
Social network structures are typically explored through measuring the number of nodes, edges, communities, cliques, and the average shortest distance between nodes~\cite{freeman2004development}.
Real-world social networks, such as virtual friendship and protein interaction networks, exhibit a small-world phenomenon. The phenomenon indicates that a network graph is sparse, and each node can reach another via a small number of intermediary nodes (six on average)~\cite{sherchan2013survey}. 
The phenomenon ties networks from different domains under a common abstraction and makes common social network analysis approaches like centrality measures applicable across those domains. ~\cite{sherchan2013survey}.

\subsubsection{Centrality measures} Centrality measures estimate the influence of a node within a network~\cite{das2018study}. Centrality measures are global measures that provide a rating for each node from the point of view of the whole network. Centrality measures have been used in various research to estimate the central figures in the network, including, but not limited to, a network of researchers, criminals, and students~\cite{das2018study}. Researchers have proposed various centrality measures that may be applicable in different contexts~\cite{das2018study}.

\section{Methodology}
\label{sec:method}
The Rust ecosystem has an active culture of collaboration and use of open source packages~\cite{schueller2022evolving}.
Many security-critical projects, like popular blockchain networks, are being developed in Rust~\cite{yakovenko2018solana, wood2016polkadot}. Further, the Rust ecosystem has built multiple tooling frameworks to help the secure use of its packages~\cite{rustsecure}. Therefore, we chose to study the Rust ecosystem in this paper. 

In this section, we explain the methodology for building a developer social network for the Rust community and providing a rating for the developers in the network by aggregating five centrality measures. We also explain the developer survey we conducted to evaluate if our proposed rating reflects the community's perception. 

\begin{table}[]
    \centering
    \caption{An overview of collected data from October 2020 to October 2022 which we use to construct a social network of Rust developers.}
    \begin{tabular}{lr}
        \hline
        Crates & 1,644 \\
        Repositories & 1,088 \\
        Commits & 109,512 \\
        Reviewed Commits & 57,592 (52.6\%) \\
        Rejected Pull Requests & 2,975 \\ 
        \hline
        Developers & 6,949 \\
        Authors & 6,616 (95.2\%) \\
        Reviewers & 2,891 (41.6\%) \\
        \hline
        %Collaborations & 228,995 \\
        Relationships & 26,448 \\
        %File Co-edition Collaborations & 166,675 (72.8\%) \\
        File Co-edition Relationships  & 18,461 (69.8\%)\\
        %Author-Reviewer Collaborations & 62,320 (27.2\%) \\
        Author-Reviewer Relationships & 14,363 (54.3\%) \\
        \hline
    \end{tabular}
    \label{tab:dataset}
\end{table}
\subsection{DSN Construction}
\label{dsn}

In this subsection, we explain our data collection and social network construction methodology.

\begin{table*}[]
    \centering
    \caption{Centrality measures used in this study to rate developers in the Rust community}
    \begin{tabular}{p{3cm}|p{5cm}|p{5cm}}
    \hline
        \textbf{Metric} & \textbf{Definition} & \textbf{Interpretation}  \\
        \hline
        Degree Centrality &  The fraction of the total nodes in the network it is connected to.   & The extent of collaboration with other  developers indicates the developers' direct trust relationships in the network. \\
        \hline
        Closeness Centrality & The reciprocal of the average shortest path distance from the node to all other reachable nodes in the network. & How quickly any developer can reach another developer may estimate the developer's influence within the network.\\
        \hline
        Betweenness Centrality  & The fraction of all pairs in the network whose shortest paths pass through that node.
        & The developers who keep the network connected can be perceived as more influential. \\
        \hline
        EigenVector Centrality & Eigenvector centrality computes the centrality of a node based on the centrality of its neighbors.  & The collaborators of a trusted developer may also be trusted.\\
        \hline
        PageRank & PageRank computes a ranking of the nodes in the graph based on the structure of the number and quality of the incoming links. & PageRank is used to measure the influence or trustworthiness of a node in the network.\\
        \hline
        %  Community size & . & .\\
        % \hline
        % Clustering co-efficient &  Clustering coefficient is a property of a node in a network. Roughly speaking it tells how well connected the neighborhood of the node is. & \\
        % \hline
        % Node clique number & Maximum size of a clique (strongly connected component) of a node & \\
        % \hline
    \end{tabular}
    
    \label{tab:centralitymeasures}
\end{table*}
\subsubsection{Data Collection}
We obtained the metadata for packages from Crates.io from the official data dump~\cite{cratesio}. At the time of this study, Crates.io hosted 92,231 packages. For this study, we chose the most 1,000 downloaded packages and all their dependency packages, which resulted in 1,724 packages. While ideally, the developer network should include collaboration history from all the Crates.io packages and all open source projects collaborated by the package developers, we chose to work on a subset of the packages as our data collection methodology is constrained by GitHub API rate limit. While we discuss the limitations of our chosen network boundary in Section~\ref{limitation}, prior work has found DSN metrics to be robust even with incomplete data~\cite{nia2010validity}.

Out of the 1,724 packages, we found valid GitHub repositories for 1,644 packages. We restricted our study to only GitHub repositories, as we identify distinct developers through GitHub accounts and collect code review information from GitHub. These 1,644 packages are hosted on 1,088 distinct repositories. To build the developer network, we considered code activities over a two-year period between October 2020 and October 2022. Specifically, we collected 109,512 commits from this period. 

For each commit, we collected the GitHub user account of the author. We consider each distinct GitHub user as an individual developer, excluding the bot accounts~\footnote{the bot accounts' usernames are suffixed by `[bot]' on GitHub}. We also determined if each commit was reviewed or not by developers other than the author. We consider a commit to be code reviewed if there is a review approval on the associated pull request on GitHub or the commit was merged into the codebase by a different developer~\cite{imtiaz2022phantom} We identified 52.6\% (57,592) of the studied commits to have been code reviewed. We also collected data on the rejected pull requests. If a non-merged pull request has a review from or was closed by a different developer, we consider that as a collaboration between the author and reviewer. In total, we obtained 2,975 pull requests rejected by a reviewer.

Overall, we found 6,949 distinct developers in our data set who are authors or reviewers of the studied commits.
Of these,  6,616  developers have authored at least one commit. Similarly, 2,891 developers have reviewed a commit at least once. 
We construct a social network over these 6,949 developers in the following subsection.
% On the other hand, 4,058 authors never acted as a reviewer, while 333 reviewers do not have a commit authored within our data set. 
Table~\ref{tab:dataset} provides an overview of the study dataset.

\subsubsection{Developer Social Network (DSN) Construction}
% \begin{figure}
%     \centering
%     \includegraphics[scale=0.5]{degree-dist.pdf}
%     \caption{The power-law distribution of node linkages (log-log scale) in the social network indicates a small world phenomenon.}
%     \label{fig:nodelinkage}
% \end{figure}

A DSN is a graph data structure where the nodes represent developers, and the edges may represent communication and collaboration among them~\cite{herbold2021systematic}. We focus on developer collaboration in this study with the rationale that collaboration can be mined directly from the repository history, unlike communication which can spread across various channels, such as emails and Slack. We use two metrics to capture developer collaborations~\cite{kerzazi2016can, meneely2011socio}:

\begin{enumerate}
    \item \textbf{File Co-edition Collaboration:} If two developers work on the same file within a 30-day window, we consider that activity to indicate collaboration between them. This metric was validated by Meneely et al.~\cite{meneely2011socio}.

    \item \textbf{Author-Reviewer Collaboration:}  If a developer has reviewed code changes submitted by another, we consider that activity to indicate collaboration between the author and the reviewer. 
    %The review outcome may be either approval or rejection. 
\end{enumerate}

We identified 166,675 instances where two developers had made changes to the same file within a 30-day period, involving a relationship between 18,461 developer pairs. We also identified 62,320 cases where a developer had reviewed code from another developer, involving 14,363 relationships. Combining both types of relationships (the same developer pair can have both types of relationships), we have edges between distinct 28,448 developer pairs. Further, we set edge weights in the network as the number of collaborations following prior work~\cite{kerzazi2016can, meneely2011socio}. Finally, we constructed an undirected, weighted graph based on our collected data using the Python package, \textit{networkx}~\cite{hagberg2020networkx}.

\subsection{Centrality Measures for Developer Rating}
\label{sec:centrality}
 Centrality measures are a common approach to estimating subjective concepts like trust, reputation, and influence~\cite{bosu2014identifying, asim2019trust, ceolin2017social, meo2017using, zahi2020improved, csimcsek2020combined}. Overall, centrality measures indicate who are the central actors in a network and, thus, can be used to rank the actors within the network~\cite{riquelme2018centrality, cadini2009using}. In DSN literature, Bosu et al.~\cite{bosu2014impact} have studied the impact of developer reputation on code review outcomes by identifying the core developers in a network through six centrality measures. We incorporate five of those six metrics in our work, as these were also used in a trust model proposed by Asim et al.~\cite{asim2019trust}. One metric we excluded from Bosu et al.'s~\cite{bosu2014impact} work is eccentricity, which only works for graphs with a single connected component, and is unsuitable in our context. Table~\ref{tab:centralitymeasures} explains the five metrics and provides our rationale behind using them in our study context. 

We compute the five metrics using \textit{networkx} package that considers the weights of the edges. We then aggregate them to avoid bias from any single measure~\cite{bosu2014identifying, csimcsek2020combined}. We take the normalized value of each of the five centrality measures listed in Table~\ref{tab:centralitymeasures}, then calculate the normalized sum of those values to get a final rating between 0 and 1. All five centrality metrics are ordinal attributes, so their aggregated rating will also be ordinal. Therefore, we use the rating to rank all the developers in our constructed network, where a higher rating indicates higher centrality for a developer. Our approach of a simple additive weighting of normalized scores is commonly followed when combining multiple metrics for ranking~\cite{tofallis2014add, shabrina2022investigating, cody2018investigation}. 

\label{devsurvey}

  \begin{table*}[]
    \centering
    \caption{Survey Questions}
    \begin{tabular}{|ll|l|}
    \hline
    \multicolumn{2}{|l|}{\textbf{Questions}}& \textbf{Options}\\
    \hline
    \textbf{Q(a)} & \makecell[l]{Please rate the below statements based on how you review \\the incoming upstream code changes in your dependency packages \\(when adding or updating a package).\\
    --- The question presents six statements which are listed in Figure~\ref{fig:q5} \\when discussing the findings.} & \makecell[l]{Always, Often, Sometimes, \\Rarely, Never}\\
    \hline
        \textbf{Q(b)} & \makecell[l]{ In the Rust package ecosystem context, choose one of the five options for \\the below-listed GitHub users.\\
        --- The respondent is presented with a list of ten GitHub users\\who are sampled separately for each respondent. \\The sampling strategy is explained in Section ~\ref{sec:survey}.
        } & \makecell[l]{
        1. I have never heard of this person before.\\2. I recognize this name, but I don't know\\ much about them.\\3. I know this person, but I don't know anyone\\ who has worked with them.\\4. I know this person, and I have worked with \\people who have worked with them.\\5. I have directly worked with this person.
        }\\
        \hline
         \textbf{Q(c)} & \makecell[l]{ When adding or updating a package, how carefully would you review\\ upstream code changes coming from the below-listed GitHub users?\\
Please choose one of the four options based on your knowledge/association\\ with them. Note that the code changes are within the dependency packages \\of your project. For example, when updating a package, how do you review the \\new changes that come with the update?\\
        --- The respondent is presented with the same list of\\ ten GitHub users from Q(b).} & \makecell[l]{ 
        1. I would not include code from \\this person in my project. \\
        2. I would review each line of upstream \\code change coming from this person. \\3. I would skim through the upstream code \\changes coming from this person.\\4. I do not feel the need to review the upstream \\code changes coming from this person.} \\
\hline
    
    \end{tabular}
    \label{tab:survey_questions}
\end{table*}
\subsection{Developer Survey}
\label{sec:survey}
To evaluate our ranking approach (RQ3), we want to understand if developers with higher centrality ratings in our constructed network are perceived to be more trusted by the community in terms of their code facing less scrutiny from the downstream projects. To that end, we conduct a developer survey to 
(i) understand if authors' and reviewers' identities impact how carefully the respondents review upstream code, and (ii) collect data on the level of scrutiny the respondents would put in reviewing upstream code coming from different developers in the network. Finally, we use the data indicating the level of scrutiny assigned to different developers as a proxy measure of trust and conduct a correlation test with the developers' centrality ratings. We sent the survey to a subset of the developers in the network who have directly collaborated with at least five others in the network, with the rationale that the developers who only had a few interactions in the community may not be able to provide informed answers about other developers.

Our survey contained three required questions. Table~\ref{tab:survey_questions} shows the survey questions and corresponding Likert-scale options. We detail each of the  questions below:
% \begin{itemize}

    \indent \textbf{Q(a)} 
    The motivation behind Q(a) is first to understand how often developers review upstream code changes before merging them into their codebase. We also want to know if the authors' and reviewers' identities impact their review process. For example, we ask if the authors' and reviewers' reputations within the Rust community and the recipients' familiarity with their work impact their review process. Q(a) helps us to establish the motivation that a ranking system for Rust developers will be helpful for the downstream project developers to review upstream code. 

    \indent  \textbf{Q(b)} The motivation behind Q(b) is to understand if our constructed network reflects the true collaboration relationships between the studied developers. The responses of Q(b) help us validate the network we used to calculate the centrality measures and ratings for each developer. To this end, we sample ten developers from the network independently for each respondent and ask them how well they know these developers. 

    We sample ten developers for each respondent in the following ways: (i) five developers from the direct collaborators of the recipient as indicated by an edge in our constructed network, and (ii) five developers from the rest of the network with whom the recipient does not have an edge with. Within these two groups (direct collaborators and others), we sample three from the top 50 according to the ranking we derive from centrality measures described in Section \ref{sec:centrality}, and two from the rest of that group. We chose this sampling approach, so the participants have a higher chance of being familiar with at least one of the developers. Further, this approach ensures that the survey recipients have top, average, and low-ranked developers to provide their opinion.

    \indent  \textbf{Q(c):} The motivation behind Q(c) is to evaluate if our proposed ranking approach reflects developer perception in the community. We present the survey recipient with the same ten developers from Q(b) and ask how the recipient would review upstream code changes coming from them. We provide four Likert-scale options to indicate the level of scrutiny the recipient would put into upstream code from each developer. Using the responses to Q(c), we statistically test if the recipients' answers correlate with the developers' centrality rating. 
% \end{itemize}

 \begin{table}[]
    \centering
    \caption{Structural overview of the social network of Rust developers}
    \begin{tabular}{lr}
    \hline
        Nodes & 6,949 \\
        Edges & 26,448 \\
        Components & 132 \\
        Isolates & 111 \\
        Network density & 0.001 \\
        Avg. clustering coefficient & 0.398\\
        Total communities & 164 \\
        \makecell[l]{Communities with more than a hundred nodes} & 21 \\
        \hline
        \multicolumn{2}{c}{Largest Component} \\
        \hline
         Nodes & 6,789 (97.7\%) \\
         Edges &  26,417 \\
         Avg. shortest path length & 3.674 \\
         \hline

    \hline
    \end{tabular}
    
    \label{tab:dsn}
\end{table}

\section{RQ1 Findings}
\label{sec:rq1}
\textbf{To what extent are developers from the most downloaded Rust packages interconnected through collaboration?}
Our constructed network graph consists of 6,949 nodes (developers) and 26,448 edges (collaborations). The largest connected component\footnote{A connected component in a graph is a group of nodes where every two nodes have a direct or indirect path (via other nodes) between them.} in the network consists of 6,789 nodes (97.7\%), showing the majority of the studied developers are interconnected. The rest of the developers are isolated or connected within a small group of collaborators. The density~\footnote{The ratio of existing edges to all possible edges.} of our network is 0.001, and the average clustering coefficient~\footnote{The clustering coefficient of a node is a measure of the extent to which its neighbors are also connected to each other.} of the nodes is 0.4, which are typical of real-world social networks~\cite{walsh1999search}. The low density and clustering coefficient values indicate the graph consists of many tightly-knit clusters. 

\begin{figure}
    \centering
    \includegraphics[scale =0.55]{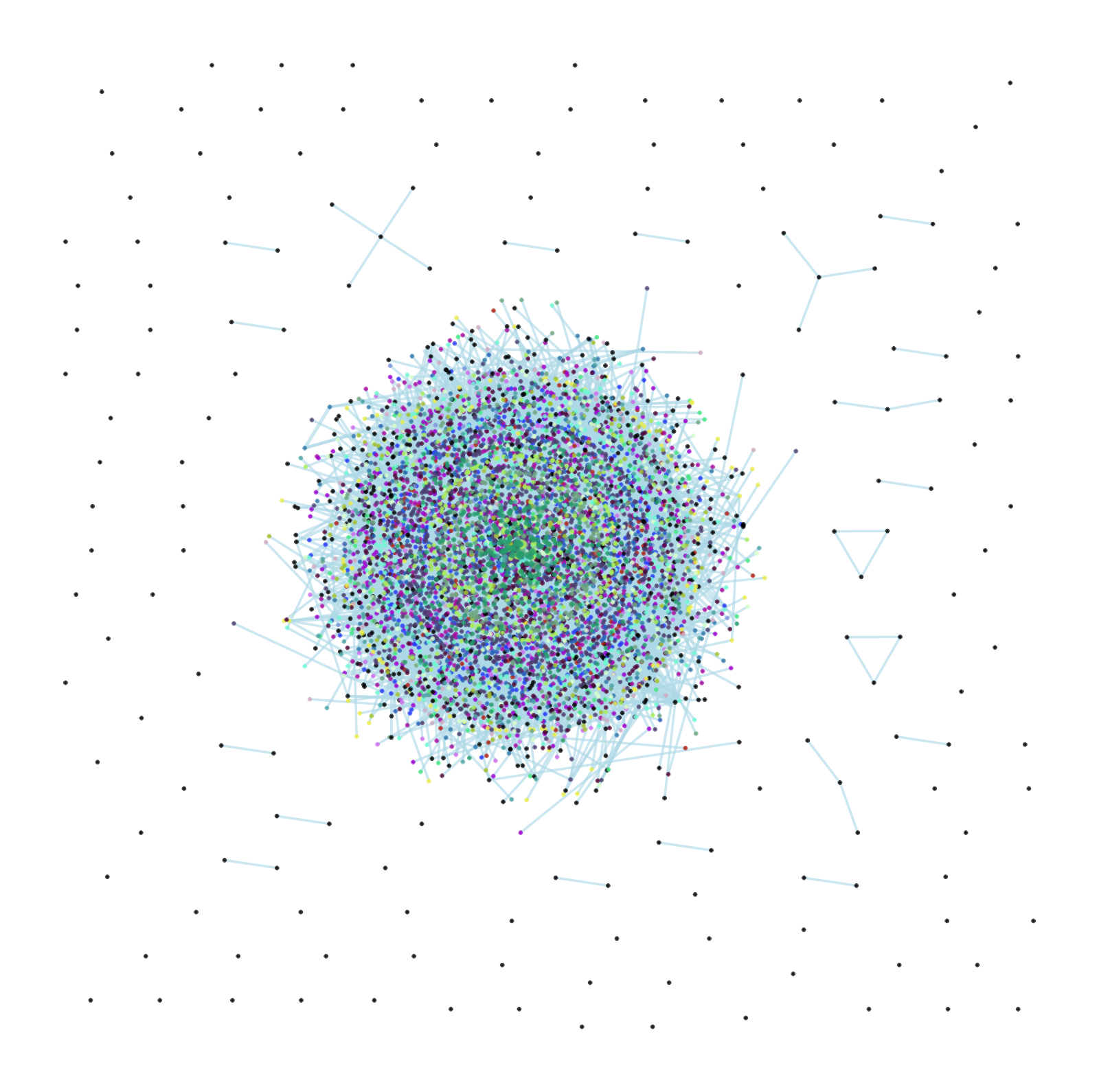}
    \caption{Social network graph of Rust developers}
    \label{fig:dsn}
\end{figure}

Further, the average shortest path length in the largest connected component is 3.7 meaning any two developers are connected via approximately 4 other developers, on average, indicating a small-world phenomenon~\cite{walsh1999search, sherchan2013survey}. 
% The small-world phenomenon establishes the applicability of common social network analysis methods, e.g., centrality measures, over our constructed network~\cite{watts1998collective, wasserman1994social}.
Further, we found 164 communities~\footnote{A subset of nodes within the graph such that connections between the nodes are denser than connections with the rest of the network.} using the Louvain community detection algorithm~\cite{que2015scalable}. However, most developers are members of 21 communities that consist of at least 100 developers. The community structure further establishes the interconnected nature of the Rust developer network. Table~\ref{tab:dsn} summarizes the properties of the network. 
Figure~\ref{fig:dsn} visualizes the network, where the nodes in the communities with larger than 100 developers are assigned distinct colors. 

Overall, this RQ reveals the properties of our constructed Rust developer network. We find Rust developers to be interconnected within tightly-knit clusters. The network shows characteristics of a small world, which establishes the ground for centrality analysis~\cite{watts1998collective, wasserman1994social}, presented in the following Section~\ref{sec:rq2}. 

\section{RQ2 Findings}
\label{sec:rq2}
\begin{figure}
    \centering    \includegraphics[scale=0.3]{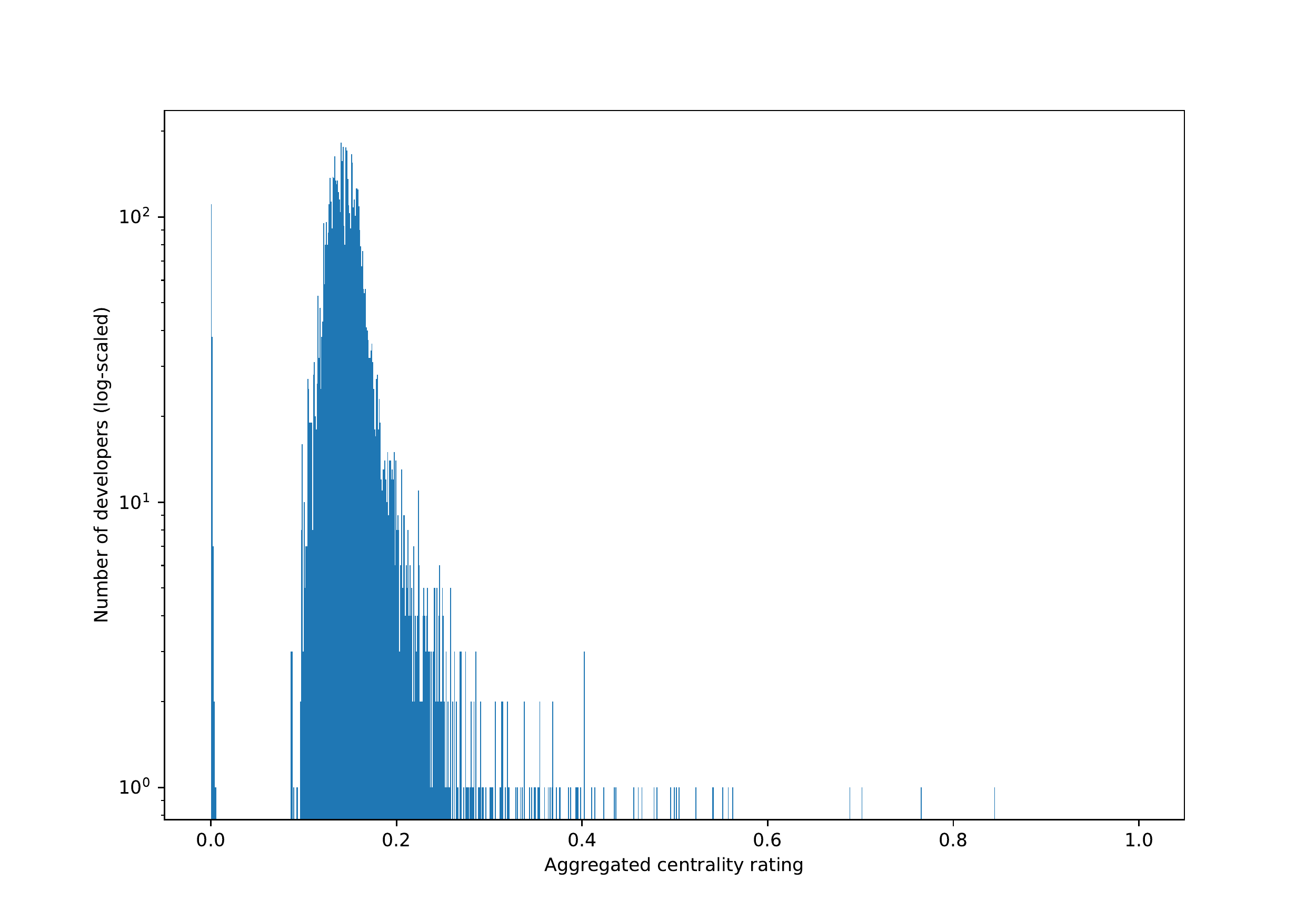}
    \caption{Histogram of aggregated centrality rating of Rust developers}
    \label{fig:centralized}
\end{figure}

\textbf{What is the distribution of the developer ratings based on their network centrality?}
As explained in Section~\ref{sec:centrality}, we aggregate five centrality measures to provide a single rating to each developer in the network. Figure~\ref{fig:centralized} shows a histogram of the aggregated rating for the developers in our network. The histogram shows a skewed distribution of the aggregated centrality ratings. We find 5.3\% of the developers to have a rating above 0.2, indicating their core positions in the network. Conversely, 2.6\% of the developers have a rating below 0.1, indicating their peripheral position. The ratings for the rest of the 92.1\% of the developers are distributed between 0.1 and 0.2. While the aggregated rating offers a granular ordering among the developers, the skewed distribution can be visually interpreted to categorize them into high (>0.2), low (<0.1), and average (0.1-0.2) ranks based on their network centrality.
 
\begin{figure*}
    \centering
    \subfloat[\label{fig:commit}]{} 
    {{\includegraphics[scale=0.3]{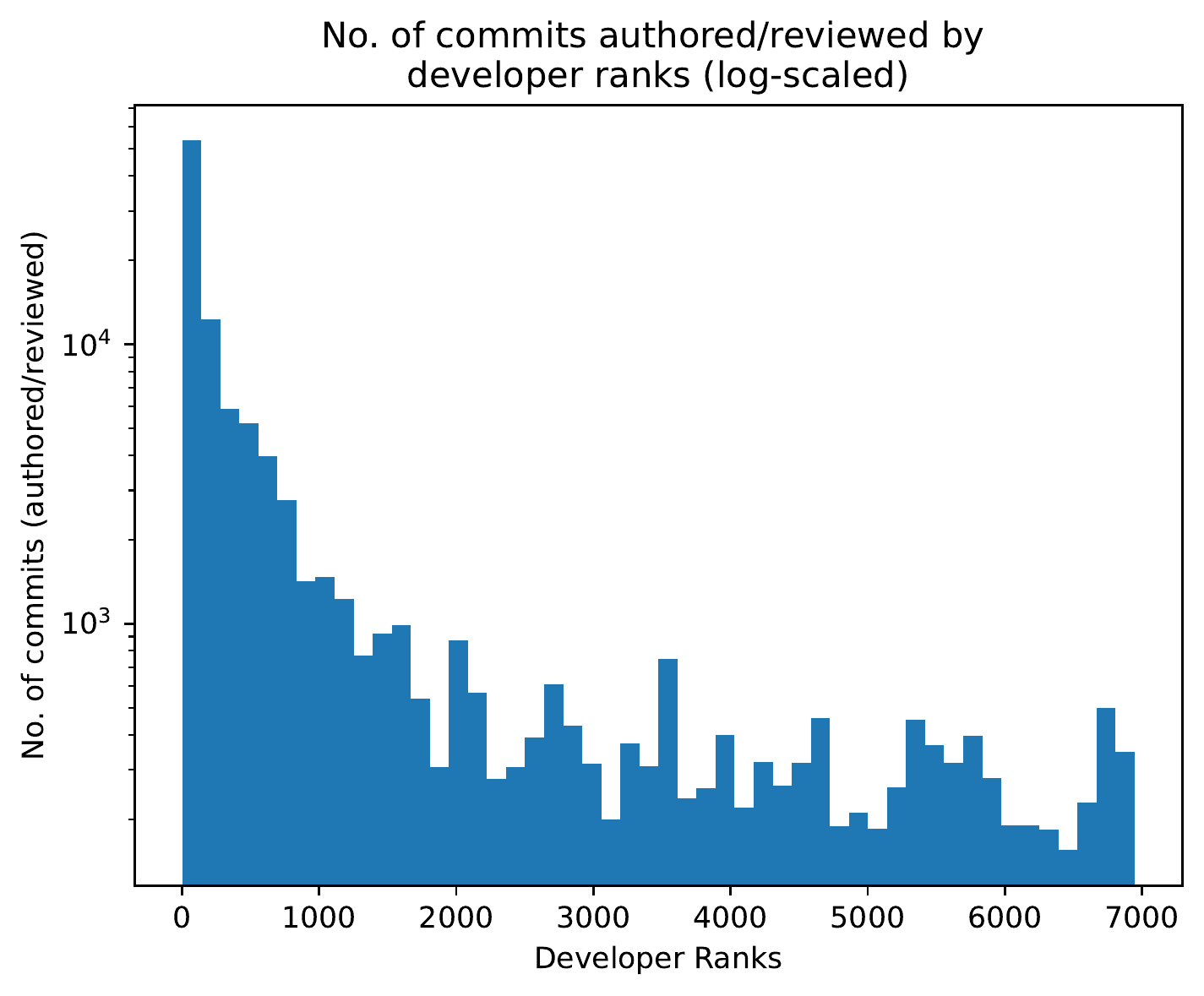} }}
    \subfloat[\label{fig:repos}]{}
    {{\includegraphics[scale=0.3]{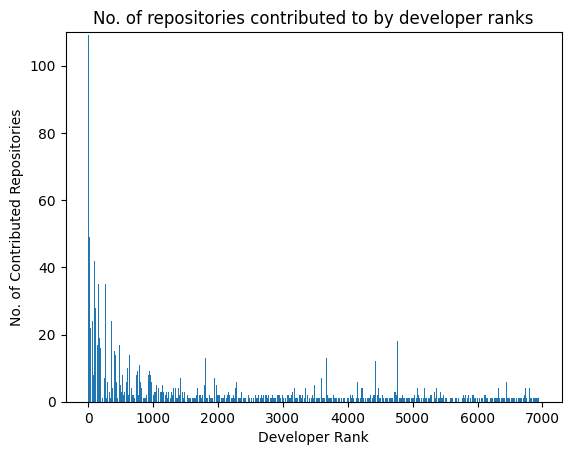} }
    \subfloat[\label{fig:collab}]{}
    {{\includegraphics[scale=0.3]{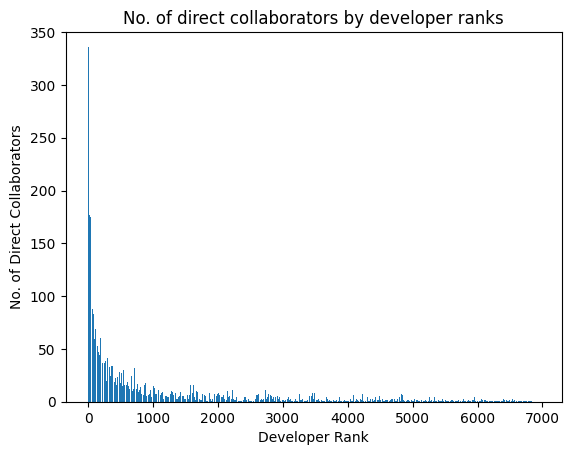} }}
    }
    \caption{Bar chart for developer rank vs. (a) no. of commits contributed either as an author or a reviewer, (b) no. of repositories the developers contributed to, and (c) no. of direct collaborators. In the x-axis, developers are sorted in descending order based on their centrality ratings, i.e., ranked first to last, where the first-ranked developer has the highest centrality.}
    % in the x-axis, developers are sorted in descending order based on their centrality ratings, i.e., ranked from high to low where a lower rank has higher centrality rating than that of a higher rank
    \label{fig:rq2}
\end{figure*}

 % \begin{figure}
 %     \centering
 %     \includegraphics[scale=0.4]{rq2.pdf}
 %     \caption{Frequency plot for no. of commits vs the rank (per network centrality) of the developer who authored/reviewed them.}
 %     \label{fig:powerlaw}
 % \end{figure}
% We use this aggregated rating to rank the developers in the community. Based on the ranking, the GitHub usernames of the top 25 developers in the community are:
% \textit{djc,
%  alexcrichton,
%  flip1995,
%  taiki-e,
%  dtolnay,
%  davidpdrsn,
%  epage,
%  camsteffen,
%  Jarcho,
%  Darksonn,
%  JohnTitor,
%  xFrednet,
%  seanmonstar,
%  ehuss,
%  GuillaumeGomez,
%  robjtede,
%  jplatte,
%  hawkw,
%  Amanieu,
%  tarcieri,
%  joshtriplett,
%  jyn514,
%  smoelius,
%  Manishearth,
%  Thomasdezeeuw.}

 % Based on the interpretation of the 5 centrality metrics that are aggregated in the centrality ratings, the rating should capture the degree of all direct or indirect collaboration (as author or reviewer) of the developers. Higher collaboration refers to higher centrality and thus, higher rating. Our investigation on the collaborators of the commits in our dataset validated that the centrality rating correctly represents the degree of collaboration of developers with others. 

We investigated developers' characteristics in terms of the relationship between their centrality ratings and (a) the number of commits they contributed either as an author or a reviewer; (b) the number of different repositories they contributed to; and (c)  the number of developers they directly collaborated with.  Figure~\ref{fig:rq2} shows a bar chart on each characteristic across developers from different centrality ratings (in the x-axis, developers are sorted in descending order based on their centrality ratings, i.e., ranked first to last, where the first ranked developer has the highest centrality.). Our results show that developers with high centrality ratings are also the ones to participate (as authors or reviewers) in the most number of commits and repositories and have collaboration with the most number of developers in the network. 

When looking at the commits specifically in Figure~\ref{fig:commit}, our dataset has 51,154 unreviewed commits from 883 distinct authors. Our results show that the top 10\% of developers based on network centrality have authored 78.2\% of these unreviewed commits. For 57,592  reviewed commits (52.6\% of the total), more than one developer is involved in the commit. Considering both authors and reviewers of these commits,  the top 10\% developers were involved in 89.7\% of these reviewed commits.  Overall, our results show that the top 10\% of the developers were involved in 84.2\% of the commits in our dataset, either as an author or a reviewer. However, note that if we scale our social network to all 92K packages hosted on Crates.io, a similar pattern between the centrality ratings and the number of commits authored/reviewed may not be found. If a developer has published many packages that are not widely used and has little collaboration with others in the network, that developer will receive a low centrality rating despite having a high code contribution.

Prior work has shown that software projects have ``hero'' developers who are core to the project~\cite{majumder2019software, agrawal2018we}. We also find the phenomenon to exist in the Rust ecosystem, as evidenced by the skewed distribution of the centrality ratings of the developers. The phenomenon works as supporting evidence that social network centrality ratings can capture the ``hero'' developers in the Rust ecosystem, who we hypothesize should be more trusted by the community. We test our hypothesis by investigating RQ3 through a survey.

%  \begin{tcolorbox}
%     We find that top developers based on network centrality were involved in most of the commits in our dataset, with top 10\% developers being involved in 84.2\% of the commits.
% \end{tcolorbox}

\begin{figure*}
    \centering
    \includegraphics[scale=0.5]{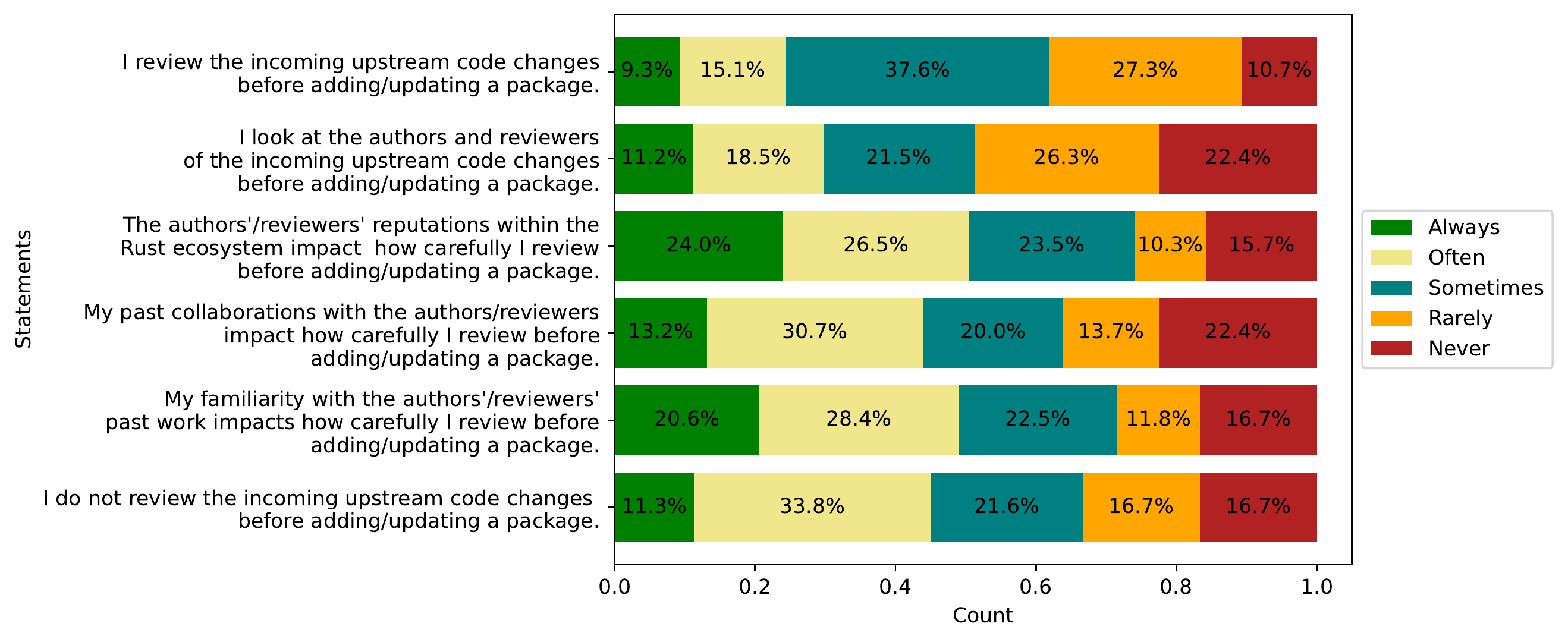}
    \caption{Likert scale responses on how developer identity impacts review process of upstream code.}
    \label{fig:q5}
\end{figure*}

\section{RQ3 Findings}
\label{sec:rq3}
\textbf{Are developers with higher centrality ratings likely to have their code accepted with lesser scrutiny by the downstream project developers?}
We investigate RQ3 through responses from our developer survey, consisting of three questions in Table~\ref{tab:survey_questions}. The questions help us to investigate if (i) authors' and reviewers' identities impact how carefully developers review upstream code; (ii) if our constructed DSN accurately captures the collaboration relationship between developers; and (iii) if developers with higher centrality ratings are perceived to be more trusted by the community, in terms of how much their code can get scrutinized by the downstream developers.

We sent our survey to 1,995 developers from our network who collaborated with at least five others according to our collected data. We received 206 responses, with a response rate of 10.3\%. We present findings from the responses in this section.

\subsection{Q(a) responses}

In Q(a), we investigate whether the authors' and reviewers' identities impact how carefully downstream project developers review upstream code changes (i.e., review scrutiny) through a Likert-scale matrix question of six statements. Figure~\ref{fig:q5} lists the six statements and the response distribution. 

The survey participants have responded that they are unlikely to review the upstream code changes before merging them into their projects. Only 24.4\% of the respondents answered that they (always or often) review the upstream changes. Similarly, only 27.7\% of the respondents answered that they (always or often) look at the authors/reviewers of the upstream changes before adding/updating a package. In the survey, we also gave options to provide additional comments. In these open-ended responses, the respondents indicated that they are most likely to review a package before adding it for the first time and may not review the changes in subsequent updates. Further, respondents have indicated they may only read the \textit{changelog} or \textit{release notes}~\footnote{a natural text documentation of the changes in a new version} during an update and not the actual code changes. The respondents have also mentioned that the context of their projects, e.g., how security-critical the project is, and the nature of the upstream package, e.g., if the package code is internet-facing, impact if they review the changes in a new update.

While the survey respondents are less likely to review all upstream code changes, they indicated that the authors' and reviewers' reputations in the Rust community impact their decision process before adding or updating a package. \textbf{More than half of the respondents (50.5\%) suggested that the reputations of the upstream developers impact (always or often) how carefully they review upstream changes.} In contrast, only 25.9\% of the respondents indicated that the developer's reputation rarely or never impacts their review process. Similarly, respondents indicated that familiarity with the upstream developers' work (49.0\% of the responses were always or often) and past collaborations (43.9\% of the responses were always or often) impact their review process.

Overall, respondents noted that once an initial trust is established, for example, through a review while adding a package, they are less likely to review the code changes in each subsequent update. We also find that, for more than half the respondents, the identity of the authors and reviewers, e.g., their reputation in the Rust community, can play a role in how carefully they would review upstream changes.

\subsection{Q(b) responses}
% \textbf{Our constructed DSN may overestimate the direct collaboration relationships between developers.} In Q(b), we ask the survey participants how familiar they are with the given ten developers (explained in Section~\ref{devsurvey}). 

An objective behind asking this question was to investigate if our constructed social network accurately reflects developer collaborations in real life. 
Considering ten responses for ten given developers from each respondent in Q(b), we have 2,060 total data points. In 989 of these cases, our constructed DSN has an edge between the respondent and the given developer. However, only in 185 of these cases (18.7\%), the respondent answered that they directly worked with the given developer. In another 424 cases (42.9\%), the respondent indicated some familiarity with the given developers (any option other than \textit{I have never heard of this person before}). In the remaining 380 cases (38.4\%), the respondent stated they had never heard of the person before.

While the file co-edition relationship may not ensure the two developers are, indeed, familiar with each other, we find that the survey respondents failed to recognize developers in the case of author-reviewer relationships as well. Out of the 989 cases, there are 531 cases where our DSN shows an author-reviewer relationship between the recipient and the given developer. However, only in 164 (30.9\%) of these cases, the respondents indicated a direct collaboration, while in another 328 (61.8\%) cases, the respondents showed some familiarity. 
In the remaining 39 cases (7.3\%), the respondent stated they had never heard of the person before, even though our data shows an author-reviewer relationship between them. 
Conversely, there are 1,071 cases where our constructed DSN does not show an edge between the survey respondent and the given developer. In only 23 cases (2.1\%), however, the respondent indicated they had worked with the developer.

Given the distributed nature of open-source development, a developer may not be familiar with all others working on the same project at the same time. Further, a developer may not remember every collaborator's name as well. 
Overall, \textbf{the survey responses show that our constructed DSN may overestimate collaboration relationships between two developers.}

\subsection{Q(c) responses}
We hypothesize that developers with higher centrality ratings are more trusted in the community in terms of their code facing lesser scrutiny from the downstream project developers. We test our hypothesis through responses for Q(c). Out of the 206 respondents, 10 did not fill out an option for all the given developers. We present our analysis for the remaining 196 responses in this subsection.

\begin{figure}
    \centering
    \includegraphics[scale=0.35]{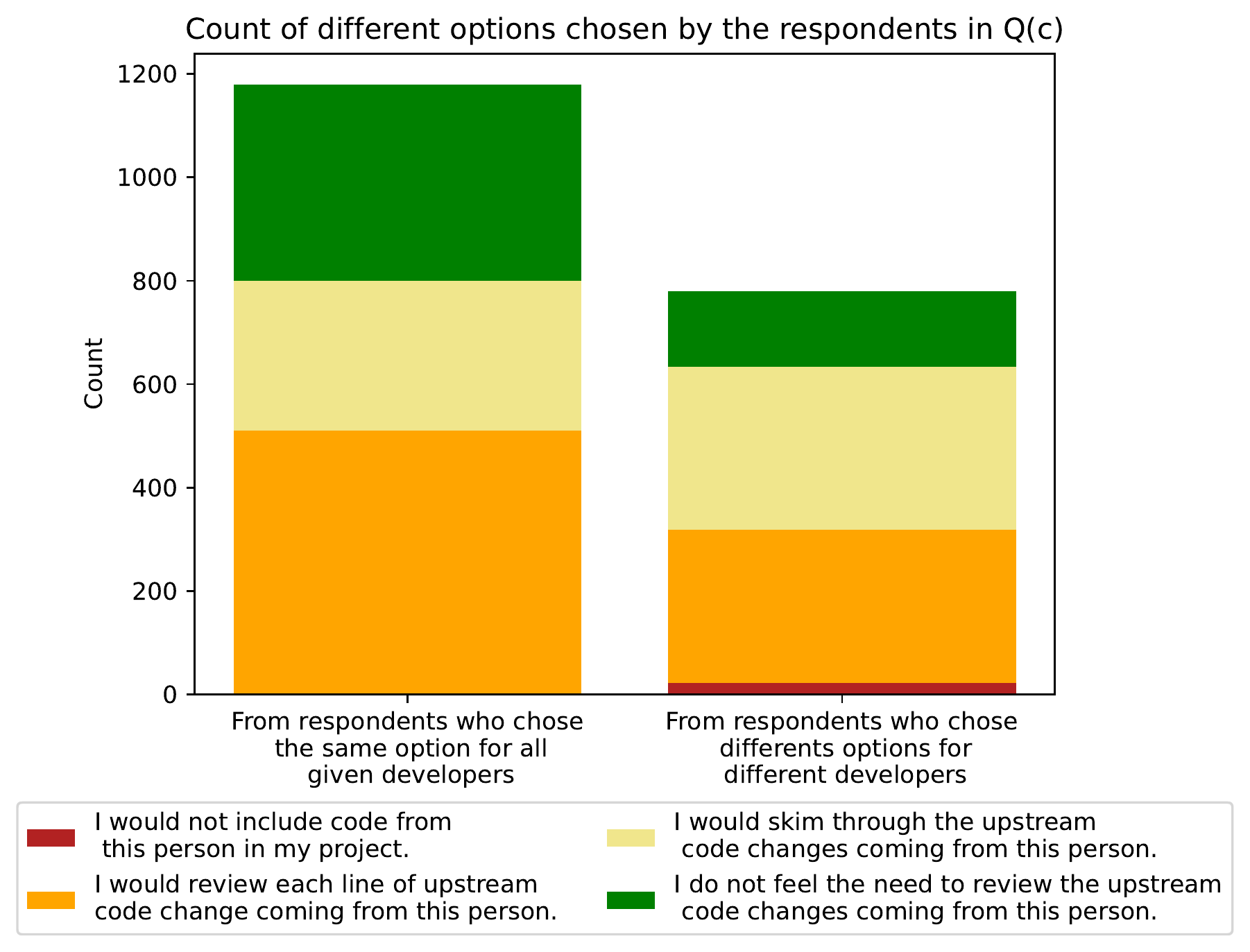}
    \caption{Count of different options chosen by the respondents in Q(c)}
    \label{fig:q4}
\end{figure}

\textbf{We find that 118 respondents (60.2\%) chose the same option for all the developers. On the other hand, 78 respondents chose different options for different developers, indicating that the developer identity may impact the level of review they employ for upstream changes. }Figure~\ref{fig:q4} shows the response distribution from both types of respondents. 

In our survey, we asked the respondents to provide an optional explanation if they chose the same option for every developer in Q(c). We received an explanation from 97 respondents. In 30 of these cases, the respondents appeared to misinterpret our question and answered based on how they would review \textit{pull requests} in their own projects, rather than changes in upstream code. Consequently, these respondents selected that they would review each line of changes. The rest of the respondents either mentioned that they do not review upstream dependency code, are not familiar enough with the given developers, or would employ a similar level of review, irrespective of who they are coming from. 

For the 78 respondents who chose a different level of review for different developers in Q(c), we measured if a developer's aggregated centrality rating correlates with the level of the review indicated by the respondents in Q(c), with 1 being the highest scrutiny and 4 being the least scrutiny. The corresponding Likert-scale options for each level of review are listed in  Figure~\ref{tab:survey_questions}.

In our analysis, we control for the fact that there can be multiple responses for the same developer coming from different respondents. To this end, we run a mixed effect regression analysis~\cite{faraway2016extending} on the 78 responses, where the respondent chose different levels of review for different developers. Mixed effect linear regression (MELR) models the association between some fixed effect variables and dependent variables when clusters or categorizations are present in the data. The regression model eliminates the impact of the clusters (that are called random effects) and represents only the correlation between the fixed effect and the dependent variables. In our model, we define the level of review indicated by the respondents for a given developer as the dependent variable and the aggregated centrality rating of the developer as the fixed effect variable. We add the developer identity as a random effect variable in the regression analysis since multiple responses for the same developer can introduce clusters in the data (780 responses are distributed over 484 distinct developers). 

\begin{figure}
    \centering
    \includegraphics[scale=0.4]{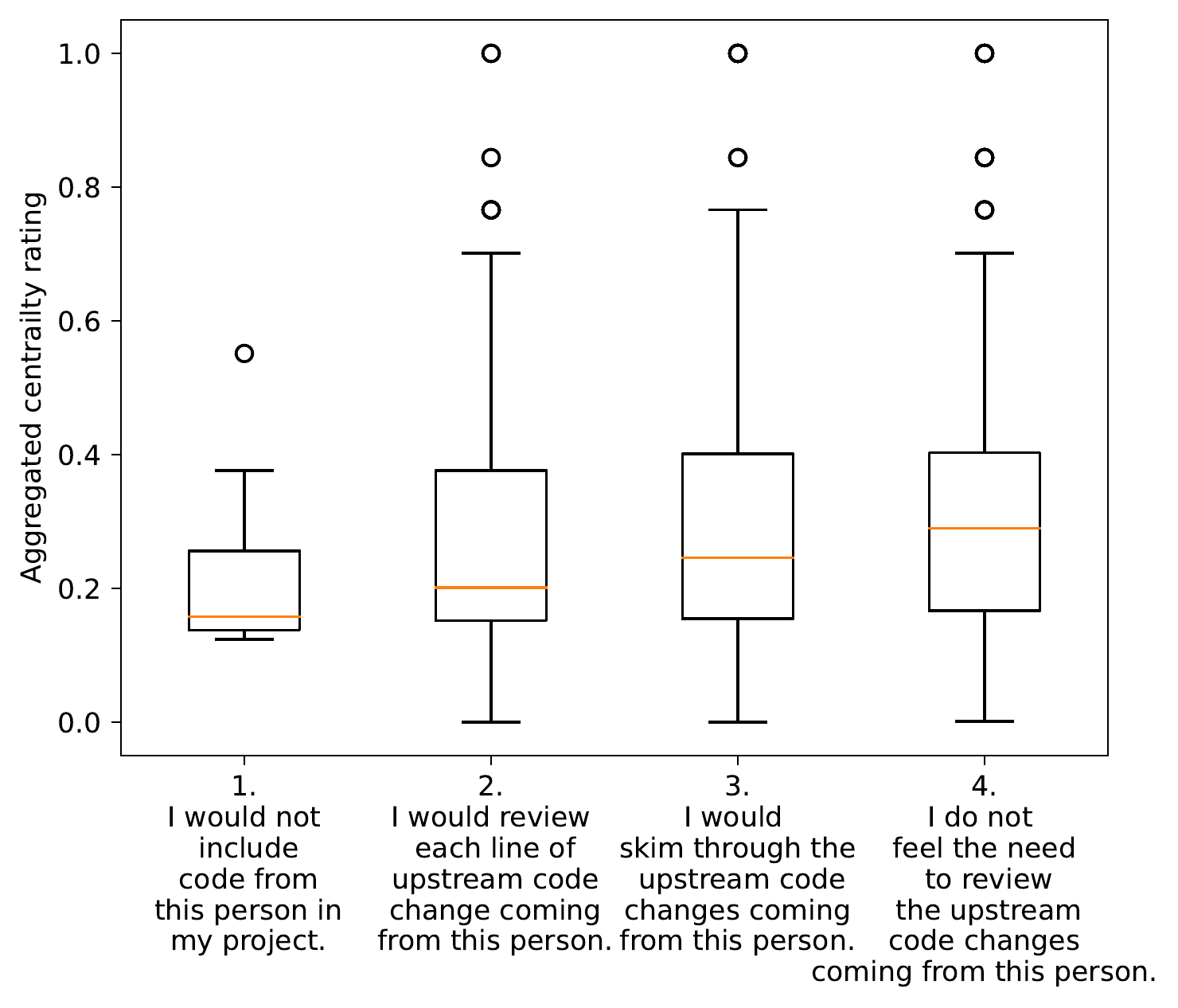}
    \caption{Box plot of aggregated centrality ratings for developers across different levels of review chosen by respondents ($N=78$) in Q(c), with 1 being the highest scrutiny and 4 being the least scrutiny.}
    \label{fig:boxplot}
\end{figure}

The p-value obtained from the MELR model shows a significant correlation between the aggregated centrality rating of a developer and the level of review or scrutiny their code may face from the downstream project developers (N=780, No. of Groups = 484, Coef. = 1.033, $p \approx 0.0$)~\footnote{We also find a significant correlation when considering responses from all 196 respondents (N=1960, No. of Groups = 991, Coef. = 0.524, $p \approx 0.0$).} — that is—\textbf{developers with higher centrality ratings are likely to be subject to less scrutiny from the downstream project developers.} We have also visualized the correlation in Figure~\ref{fig:boxplot} through a box plot of the distributions of the developers' centrality rating under each level of review, as indicated by the survey respondents. The box plots show that as the median of centrality rating decreased, the level of review scrutiny increased (from level 4 to level 1). The box plot also shows that the developers whose code the respondents chose to accept without a review (note the box plot for level 4) have a higher median centrality rating than the developers for whose code the respondents warranted comparatively more scrutiny.

Further, we investigated if the level of familiarity of the respondent with a given developer, as indicated by the responses in Q(b), correlates with the level of review or not. However, we do not find any significant correlation ($p=0.843$) in the MELR model. We also did not find the past collaboration of the respondent with the given developer to be an influential factor ($p=0.682$). Our findings indicate that centrality ratings computed from the social network can predict how carefully downstream projects review upstream changes coming from a developer and, therefore, can be used as a proxy for the trust rating of a Rust developer within the community.

\section{Discussion}
\label{sec:discussion}
In this section, we discuss the implications of our findings:

\textbf{Developer social network (DSN) at a package ecosystem level.} Prior work on DSN only focused on single large-scale projects like Linux and Firefox. In the context of a single project, a rich body of prior work showed how DSN could be leveraged for many use cases, including identifying core and peripheral developers~\cite{bosu2014identifying},
predicting defects~\cite{abreu2009developer} and post-release failures~\cite{meneely2008predicting}. Ours is the first work to show that developers at a package ecosystem level are also interconnected.  
However, we only studied a limited number of packages to construct a DSN. How the DSN structure may evolve when scaled to more packages is unknown. Similarly, popular packages from other ecosystems, like npm and PyPI, may not exhibit similar inter-connectedness as the Rust ecosystem. While our study shows promise in studying DSN at a package ecosystem level, future studies are required to understand how well the approach scales or generalizes to other ecosystems. 

\textbf{The use case of network-based centrality rating towards an identity model for open source developers.} Prior research has proposed building trust in upstream code by tracing the humans behind that code~\cite{imtiaz2022phantom}. However, in the open source, anybody can open an account on GitHub and start contributing. Therefore, only knowing which account on an online coding platform has authored/reviewed certain code may not be sufficient for downstream project developers to trust that code. In that case, the network-based centrality rating 
may be used towards the development of an identity model for open source developers. With this rating approach, one can determine not only who has authored/reviewed code in a specific package, but also estimate how reputed the developer is in the community. 

Such a rating system may also help establish trust in post-release code vetting~\cite{zimmermann2019small, imtiaz2022phantom}. Tools like \textit{cargo-crev}~\cite{cargocrev} offer a distributed code review system for developers to vet published packages so that not all downstream projects have to review the package themselves. However, establishing trust in the developers who will perform the code vetting on behalf of the entire community
is also important. Our social network-based centrality rating system can be helpful in these cases to identify the most reputed developers in a community and make crowd-based code vetting objectively trustworthy.  

\textbf{Package managers and audit tools can incorporate developer centrality ratings in their reporting of package metadata.} Package managers like Crates.io and audit tools like \textit{OSSF Scorecard}~\cite{ossfscorecard}, \textit{packj}~\cite{packj}, and \textit{Depdive}~\cite{imtiaz2022phantom} provides various analysis over a package and its updates to help downstream project developers be aware of potential security, quality, and maintainability risks. For example, these tools provide the download count, code review coverage, and the maintainers' email domain validity to help developers review a new update. These tools can also incorporate centrality ratings of the authors and reviewers of upstream changes, as our work has shown such a rating can be a helpful signal.

Developers' centrality ratings can be leveraged to identify potentially malicious activities. For example, code review coverage proposed by Imtiaz and Williams~\cite{imtiaz2022phantom} can be gamed by opening a fake account to stamp a review on malicious code pushed by a compromised developer account. Further, a group of developers may be colluding, in which case code review coverage would make little sense. Developer centrality ratings and their collaboration patterns may be leveraged to identify such potentially malicious activities and prevent future supply chain attacks~\cite{ohm2020backstabber}.

\textbf{Can we measure trust at an organizational level?} One respondent in our survey noted that measuring trust at the individual developer level, as has been done in this paper, may not scale. One alternate approach to rating individual developers is to rate organizations in the community. Many open source developers contribute on behalf of large organizations, such as Mozilla, Facebook, and Google. We can leverage social network analysis to identify the top organizations in a community, similar to how we identified the top developers. 
% The top organizations may include a large number of developers and, therefore, an attribute indicating the organization of each developer may help scale larger ecosystems like npm and PyPI better. 
Future studies may build on our work in this paper and investigate if trust based on an organizational level may be appropriate at larger scales.

\section{Limitation}
\label{limitation}

In this section, we discuss the limitations of our study:

\textbf{DSN construction.} The construction of DSN in this paper has multiple limitations: (i) we only considered packages hosted on GitHub; (ii) we only considered developers with a GitHub account; (iii) we only collected code review information based on the evidence present on GitHub; (iv) we did not consider non-code based interactions between developers, like Slack and email, and (v) relationships between upstream and downstream project developers. Further, we may not have captured all existing developer relationships due to the limited data chosen to construct the DSN. When calculating edge strength between a pair of developers, we did not differentiate between different packages, the type, and size of contributions in each commit, and the complexity of different changes. While a DSN built upon more elaborate data may have given us more insights into the community structure and the developers' centrality ratings, we rely on simple design mechanisms following prior work in the literature~\cite{herbold2021systematic}. Being the first work of its kind, our goal was to evaluate the feasibility of constructing DSN at a package ecosystem level and understand if a centrality-based rating can help estimate the trustworthiness of the developers.

\textbf{Generalizability Threat.} Our work focuses on the top 1,644 packages within the Rust ecosystem. Our work may not generalize to other package ecosystems like npm and PyPI. Further, our findings may not generalize when scaled to all 92K packages hosted on Crates.io. Additionally, we only worked with data collected from October 2020 to October 2022, which may pose a generalizability threat to our findings if the data is scaled to a more extensive period.

\textbf{Misinterpretation of survey questions and reporting bias.} When discussing the answers for Q(c) in our survey, we noticed that 30 respondents appeared to have misinterpreted our questions and answered based on how they would review pull requests in their own projects. While we were able to spot such misinterpretation through the explanation provided by the respondents, there could be other unknown misinterpretations that we are unaware of. 
We only ask about ten developers to each survey recipient, chosen through a stratified sampling approach.
Our sampling approach may introduce unknown biases in our evaluation of RQ3. 
Further, the response rate in our survey is only 10.3\%. While a similar response rate is typical in developer surveys~\cite{smith2013improving}, our findings may be subject to unknown biases based on who chose to participate in our survey and who did not.

\section{Conclusion}
\label{conclusion}
In this paper, we demonstrate a social network-based centrality rating for developers within the Rust community. The rating is aimed to reflect how much a developer's code is trusted by the community so that it can help downstream project developers prioritize their review efforts for changes in upstream packages. To develop such a rating, we construct a social network based on collaboration across the most downloaded 1,644 Rust packages. We show that developers from different packages are interconnected, where each developer is connected to another via only 4 developers on average. Building on this interconnected nature of Rust developers, we propose a global centrality rating for each developer, aggregated over five centrality measures computed from the network. Finally, we conduct a survey among the Rust developers to evaluate if our proposed rating reflects the perceived trustworthiness of developers within the community. Survey responses ($N=206$) show that the respondents are more likely to not differentiate between developers in deciding how to review upstream changes (60.2\% of the time). However, when they do differentiate, our results indicate that the code from the developers with higher centrality ratings is likely to face lesser scrutiny (as per the MELR-based correlation testing). To summarize, the social network-based centrality rating approach shown in this paper can be leveraged to estimate the trustworthiness of a developer in the community and may help downstream project developers decide what level of review new upstream changes may require.

\bibliographystyle{IEEEtran}
\bibliography{bibliography}
\end{document}